\documentclass[twocolumn]{revtex4}
\usepackage{amsfonts,amsmath,amssymb,latexsym}
\usepackage{epsfig}
\usepackage{graphics}

\begin{document}

\title{ Deviations from Wick's theorem in the canonical ensemble }
\author {K. Sch\"onhammer}
\affiliation{Institut f\"ur Theoretische Physik, Universit\"at
  G\"ottingen, Friedrich-Hund-Platz 1, D-37077 G\"ottingen}

\date{\today}

\begin{abstract}

Wick's theorem for the expectation values 
 of products of field operators for
 a system of noninteracting fermions or bosons
 plays an important role in
the perturbative approach to the quantum many body problem.
A finite temperature version holds in the framework of the grand
canonical ensemble but not for the canonical ensemble appropriate 
for systems with fixed particle number like ultracold quantum gases
in optical lattices. Here we present new formulas for 
 expectation values
of products  of field operators in the canonical
ensemble using a method 
in the spirit of Gaudin's proof of Wick's theorem for
the grand canonical case. The deviations from
Wick's theorem are examined quantitatively for two simple models of
noninteracting fermions.

\end{abstract}

\maketitle

\section{Introduction}

\noindent Properties of a large system of noninteracting fermions
or bosons in thermal equilibrium are usually described using the
grand canonical ensemble with variable particle number.
 For a system of fixed number $N$
of fermions
in a closed box this provides an excellent approximation
for large enough $N$. An exception is provided by the ideal Bose gas,
where the probability distribution for the particle number in
the lowest one-particle state fails badly in the low temperature
limit as the comparison for the appropriate result within the
canonical ensemble shows \cite{Mullin, Scully}.

In a recent publication a new formula for the $n$-particle density matrices 
in the canonical ensemble was presented \cite{KTTK}. In contrast to 
Wick's theorem for the grand canonical ensemble the higher order   
 reduced density matrices cannot be expressed in terms of the one-particle
function.
Here a different approach in the spirit 
of  Gaudin's proof of Wick's theorem for the
grand canonical ensemble \cite{Gaudin,FW} is presented.

In section II  earlier results for expectation values 
of occupation numbers and products of them are discussed.
In contrast to the grand canonical ensemble no simple factorization
of the expectation values of products of occupation number operators
occurs when the canonical ensemble is used. The formulas presented
earlier involve a summation which involves all partition functions for
particle number from one to $N$, which are numerically rather unstable
for large values of $N$. 

 In section III 
 new formulas are derived
 in which expectation values of $m$-particle operators
are expressed in terms of the mean occupation numbers.
For simple models like
 one-dimensional fermions in a harmonic trap
and a zero bandwidth semiconductor
explicit results are presented and the deviations from Wick's
theorem are elucidated quantitatively in section IV. Analytical expressions for
the deviation from Wick's theorem are presented.

\section{Known results for noninteracting fermions 
in  the canonical and grand canonical  ensemble}

An $N$-particle system with Hamiltonian $H_N$ in thermal equilibrium
at temperature $T$ is described by the canonical statistical operator
\begin{equation}
\rho_c=e^{-\beta H_N}/Z_N~,~~~~Z_N=Tr_Ne^{-\beta H_N}~,
\end{equation}
where $\beta=1/(k_BT)$ with $k_B$  Boltzmann's constant and
$Tr_N$ the trace over the $N$-particle Hilbert space.
The expectation value of an observable $A$ is given by
\begin{equation}
\langle A\rangle_c=Tr_NAe^{-\beta H_N}/Z_N~.
\end{equation}

In this paper systems of $N$ noninteracting fermions 
with $H_N=\sum_{\alpha=1}^Nh_\alpha$ are considered. 
The eigenstates of such a system can
be expressed as a determinant of the single particle eigenstates
obeying the single particle Schr\"odinger
equation $h|\epsilon_i\rangle=\epsilon_i|\epsilon_i\rangle$. A
convenient way to express the $N$-particle eigenstates is by the list
of occupation numbers $\{n\}$ of these one-particle states leading to
\begin{equation}
H_N|\{n\} \rangle_N= \sum_i \epsilon_i 
 n_i  |\{n\} \rangle_N  
\end{equation}
 with $\sum_i n_i=N$.
For fermions the occupation numbers $n_i$ can take the values $0$ and $1$. 
 The canonical partition
function is given by
 \begin{equation}
\label{pf} 
Z_N(\beta)= \sum_{\{n\}} e^{-\beta  \sum_j   \epsilon_j n_j }\delta_{N,\sum_j n_j}
\end{equation}
The Kronecker delta restricting the occupation number sums makes a closed 
evaluation of $Z_N$ generally difficult. Therefore for large particle
number $N$ the grand canonical statistical operator $\rho_{gc}$ 
with varying particle number is often used.
It acts in Fock space which is the direct sum over all $N$ from zero to
infinity of the Hilbert spaces of totally antisymmetric $N$-particle states.
In this context it is appropriate to use second quantization by introducing the
 creation and annihilation operators $c^\dagger_i$ and
$c_i$ of the orthonormal one-particle state $|i\rangle$
obeying the anticommutation rules $[c_i,c_j]_+=0$ and
$[c_i,c^\dagger_j]_+=\delta_{ij}$. 
For the
noninteracting systems treated in this paper
 the  Hamiltonian $ H_0$  reads
\begin{equation}
\label{H0}
 H_0=\sum_i \epsilon_ic^\dagger_i c_i=\sum_i \epsilon_in_i
\end{equation}
if $c^\dagger_i$ creates a fermion
in the energy eigenstate $|\epsilon_i\rangle$.
The $n_i=c^\dagger_i c_i$ are the occupation number operators.
The corresponding grand canonical statistical operator reads
\begin{equation}
\rho_{gc}^{(0)}=e^{-\beta \tilde H_0}/Z_F~,~~~~Z_F=Tr_Fe^{-\beta  \tilde H_0}~,
\end{equation} 
where $\tilde H_0\equiv H_0-\mu {\cal N}$ 
with ${\cal N}=\sum_in_i$ the particle number operator
expressed in terms of the occupation number operators
and $\mu$ the chemical potential.
Because it simplifies the calculations
the grand canonical ensemble is often used as an approximate
description for a system with fixed particle number $N$,
 using $\langle {\cal N} \rangle_{gc} =N$ to fix the
 chemical potential. \\

In the rest of this section we discuss known results for the expectation
values $\langle n_i\rangle $ and  $\langle n_in_j\rangle$ with $i\ne j$
for both ensembles. We
 begin with the much simpler case of the grand canonical ensemble.
The statistical operator
$\rho_{gc}^{(0)} $ factorizes 
\begin{equation}
\label{factorize}
\rho_{gc}^{(0)}=\prod_ie^{-\beta \tilde \epsilon_i n_i}/z_i\equiv 
\prod_i  \rho_i^{(0)}
\end{equation} 
with $\tilde \epsilon_i=\epsilon_i-\mu$ and $z_i=1+e^{-\beta \tilde \epsilon_i}$.
This leads to $\langle n_i\rangle_{gc}= e^{-\beta \tilde \epsilon_i}/z_i$
 and one obtains
the Fermi function
\begin{equation}
\label{gc1}
\langle n_i\rangle_{gc}=\frac{1}{e^{\beta \tilde \epsilon_i}+1}
\equiv f( \tilde \epsilon_i)~.
\end{equation} 
Because of the factorization of $\rho_{gc}^{(0)} $ the factorization for
the expectation value of two different occupation number operators 
easily follows
\begin{equation}
\label{gc2}
\langle n_in_j\rangle_{gc}=\langle n_i\rangle_{gc}\langle n_j\rangle_{gc}~.
\end{equation} 
This is the simplest version of Wick's theorem. The total factorization
of a product of an arbitrary number of different occupation number operators
is obvious. It is discussed in more detail in the next section.

With Eqs. (\ref{gc1}) and  (\ref{gc2}) a simple expression for 
the mean square deviation of the total particle number can be given.
With $n_i^2=n_i$ one obtains
\begin{eqnarray}
\langle {\cal N}^2\rangle_{gc}&=&\sum_{i\ne j}\langle n_in_j\rangle_{gc}+
\sum_i\langle n_i\rangle_{gc} \\ \nonumber
&=&\sum_{i\ne j} 
\langle n_i\rangle_{gc}\langle n_j\rangle_{gc}+ \sum_i\langle n_i\rangle_{gc}
\end{eqnarray} 
leading to 
\begin{equation}
\langle {\cal N}^2\rangle_{gc}-\langle {\cal N}\rangle_{gc}^2
=\sum_i\langle n_i\rangle_{gc}(1-\langle n_i\rangle_{gc})~.
\end{equation} 
As the right hand side of this equation is of order $N$
(see section III for explicit examples), the
relative width of the particle number distribution in the grand
canonical ensemble decreases like $1/\sqrt{N}$ in the large $N$ limit.\\

For the canonical ensemble
the mean occupation numbers $\langle n_i\rangle_N$ 
\begin{equation}
\label{niocc}
 \langle n_i\rangle_{N,c}  =\frac{1}{Z_N}\sum_{\{n\}} n_i
e^{-\beta  \sum_j   \epsilon_i n_j }\delta_{N,\sum_j n_j}
\end{equation}
 were early studied by Schmidt \cite{Schmidt}.
He derived a simple relation between $\langle n_i\rangle_N $
 and $\langle n_i\rangle_{N-1}$ (we suppress the index ``$c$'' for the
 rest of this section)
\begin{equation}
\label{nrecursion}
\langle n_i\rangle_N =e^{-\beta \epsilon_i}
\frac{Z_{N-1}}{Z_N}(1- \langle n_i\rangle_{N-1} )
\end{equation}
by performing the $n_i$ sum and introducing the factor $1- n_i$
in order to return to 
the complete sum over $\{n\}$ with $N$ replaced by $N-1$ in the
Kronecker delta. In the large $N$ limit $  \langle n_i\rangle_N
\approx  \langle n_i\rangle_{N-1} $
and $Z_{N-1}/Z_N= e^{\beta(F_N-F_{N-1})}\approx e^{\beta \mu}$ with
$F_N$
the free energy and $\mu$ the chemical potential holds,
leading to the Fermi function
\begin{equation}
\label{Fermif}
  \langle n_i\rangle_N \approx
  (e^{\beta(\epsilon_i-\mu)}+1)^{-1}=f(\tilde \epsilon_i)~.
\end{equation}
For arbitrary values of $N$ the recursion relation in Eq.(\ref{nrecursion})  
can be used. With the initial value  $ \langle
n_i\rangle_0=0 $
 one obtains in the first step 
   $   \langle n_i\rangle_1 =e^{-\beta \epsilon_i}/Z_N $
and easily proves by induction
\begin{equation}
\label{nsum}
\langle n_i\rangle_N  =\frac{1}{Z_N}\sum_{k=1}^N(-1)^{k-1}e^{-\beta
  k\epsilon_i}Z_{N-k}(\beta)~.
\end{equation}
The summation over all $i$ yields $N$ on the left hand side
leading to 
\begin{equation}
\label{ZNbeta}
Z_N(\beta)= \frac{1}{N}\sum_{k=1}^N(-1)^{k-1}Z_{N-k}(\beta)Z_1(k\beta)
\end{equation}
with $Z_1(\beta)=\sum_ie^{-\beta \epsilon_i}$. There are also other
ways to derive this relation \cite{BF1}. The sums in Eqs. (\ref {nsum})
 and (\ref {ZNbeta}) 
unfortunately are numerically rather unstable for large values of $N$.
\cite{BF2} They also do not provide analytical expressions in the
limiting cases discussed in section IV. 

The procedure leading to Eq.(\ref{nrecursion}) can easily be extended
to the calculation of expectation values of products of different
occupation numbers. Replacing $n_i$  by $n_in_j$ in Eq. (\ref{niocc})  
with $i\neq j$ one obtains 
\begin{equation}
\label{ninjrecursion}
\langle
n_in_j\rangle_N=\frac{Z_{N-2}}{Z_N}e^{-\beta(\epsilon_i+\epsilon_j)}
\langle (1-n_i)(1-n_j)\rangle_{N-2}~. 
\end{equation}
The large $N$ limit can be
treated as following Eq. (\ref{nrecursion}). With the additional 
assumption $ \langle n_in_j\rangle_N \approx \langle n_in_j\rangle_{N-2}$
one obtains using Eq. (\ref{Fermif}) after elementary algebra
\begin{equation}
\langle n_in_j\rangle_N \approx [(e^{\beta(\epsilon_i-\mu)}+1)
(e^{\beta(\epsilon_j-\mu)}+1)]^{-1}
\end{equation}
i.e. the simplest version of Wick's theorem $\langle n_in_j\rangle =
\langle n_i\rangle \langle n_j\rangle$ approximately holds for
large $N$ also in the canonical ensemble.
 For arbitrary values of $N$ one again can proceed
recursively. With the starting value $\langle n_in_j\rangle_2=
e^{-\beta (\epsilon_i+\epsilon_j)}/Z_2$ one can show by induction
\begin{equation}
\label{ninj}
  \langle n_in_j\rangle_N =\frac{1}{Z_N}\sum_{k=2}^N(-1)^k
Z_{N-k}\sum_{l=1}^{k-1}e^{-\beta[k\epsilon_i+(k-l)\epsilon_j]}~.
\end{equation}
This equation also readily follows from Eq. (5b) of reference 3.
Expectation values of higher products of
occupation number operators are discussed with a new approach in
the next section.

\section{Wick's theorem and weaker forms of it}

\subsection{General remarks}

In this section we address Wick's theorem
and weaker forms of it
 in a more general setting
and consider expectation values of multiple products of creation and
annihilation operators.  
A general $m$-particle operator can be written as a linear
combination of such a
 multiple product of $m$ creation and $m$ annihilation operators \cite{FW}
\begin{equation}
\label{nparticle}
A_{k_1k_2...k_m,l_1l_2...l_m}  =
c^\dagger_{k_1}c^\dagger_{k_2}...c^\dagger_{k_m}c_{l_m}c_{l_{m-1}}...c_{l_1}~.
\end{equation}
For the case of fermions treated here all $k_i$ and all $l_j$
have to be different in order to obtain a non-zero expression.
The two-particle interaction $V$ between fermions is an important
$m=2$ example. If it is treated in the Hartree-Fock approximation
its expectation value in a system of noninteracting fermions occurs.
This is one motivation for the following. In a higher order perturbative
 treatment of a two-particle interaction expectation values of
products of operators
$A^{(2)}$ occur which can be reordered into operators of the type in 
Eq. (\ref{nparticle}). 
In the following we want to evaluate the expectation value
 \begin{equation}
\langle A_{k_1k_2...k_m,l_1l_2...l_m} \rangle=
 {\rm Tr}( A_{k_1k_2...k_m,l_1l_2...l_m}\rho) 
\end{equation}
with $\rho$ being the statistical operator for the canonical 
or the grand canonical ensemble. Performing the trace in both
cases using the occupation number states $|\{n\}\rangle_N$ it is
obvious that  
the $\{l_j\}$ in   Eq. (\ref{nparticle})
have to be a permutation of the  $\{k_i\}$ in order to obtain
a nonzero expectation value. This implies
\begin{equation}
\label{productform}
\langle A_{k_1k_2...k_m,l_1l_2...l_m}\rangle =\det(\delta^{(m)})\langle
n_{l_1}n_{l_{2}}...n_{l_{m}}\rangle
\end{equation}
where   $ \delta^{(m)}$ is the $m \times m$ matrix with
matrix elements $  \delta^{(m)}_{ij}=\delta_{k_il_j} $.
This result holds for the canonical and the grand canonical ensemble. 
For $m=1$ this equation reads $\langle
c^\dagger_k c_l \rangle =\delta_{kl}\langle n_l\rangle$.

\subsection{ The grand canonical ensemble}

As mentioned in section II the factorization of  $\rho_{gc}^{(0)}$
(see Eq.(\ref{factorize})) immediately implies 
\begin {equation}
\label{Wickproduct}
\langle P^{(m)}\rangle_{gc} \equiv \langle n_{l_1}... n_{l_m}\rangle_{gc}
=\langle n_{l_1}\rangle_{gc}... \langle 
n_{l_{m-1}}\rangle_{gc}\langle n_{l_m}\rangle_{gc}~.
\end {equation}
 Introducing  the matrix
$\langle c^\dagger c\rangle^{(m)}$ with
matrix elements $ \langle c^\dagger c\rangle^{(m)}_{ij}=
\langle c^\dagger_{k_i}c_{l_j} \rangle_{gc}$ the expectation value 
of  $A_{k_1k_2...k_m,l_1l_2...l_m}$ can be written in the two forms
\begin{eqnarray}
\label{Wicktheorem}
\langle A_{k_1k_2...k_m,l_1l_2...l_m}\rangle_{gc} &=&
\det(\delta^{(m)})\prod_{i=1}^m\langle 
n_{l_i}\rangle_{gc} \\ \nonumber
=\det( \langle c^\dagger c\rangle^{(m)} )~.
\end {eqnarray}
Due the multilinearity of the determinant the second form also holds
for arbitrary operators $c_\alpha=\sum_i
\langle \alpha|\epsilon_i\rangle c_i$ in the definition of $A$.
This a general form of Wick's theorem for fermions.\\

For the attempt to express $\langle A_{k_1k_2...k_m,l_1l_2...l_m}\rangle$
in terms of the mean occupation numbers $\langle n_i \rangle$ also for
the canonical ensemble it is useful to present an alternative
way to calculate expectation values of the type 
$ \langle \tilde Ac_i \rangle_{gc}$, where $\tilde A$ is an arbitrary
product of $m$ creation and $m-1$ annihilation operators.
 The essential steps in
Gaudin's method \cite{Gaudin,FW} to determine such
expectation values for fermions or bosons
 are to use Heisenberg type operators
 \begin{equation}
e^{\beta \tilde H_0} c_i e^{-\beta \tilde H_0}=e^{-\beta \tilde \epsilon_i} c_i
\end{equation}
and the cyclic invariance of the trace.
This leads to 
\begin{eqnarray}
\label{cyclic}
{\rm Tr}_F \tilde A c_ie^{-\beta \tilde H_0}&=& 
{\rm Tr}_F \tilde A e^{-\beta \tilde H_0} e^{\beta \tilde H_0} c_ie^{-\beta
  \tilde H_0}
 \nonumber \\
&=& e^{-\beta \tilde \epsilon_i}{\rm Tr}_F\tilde A e^{-\beta \tilde H_0}c_i
\nonumber \\
&=&e^{-\beta \tilde \epsilon_i}{\rm Tr}_F c_i\tilde A e^{-\beta \tilde H_0}~.
\end{eqnarray}
Now one can use $c_i\tilde A=\mp \tilde A c_i+[c_i, \tilde A]_\pm$,
where the upper (lower) sign is for fermions (bosons). 
Multiplication with $e^{\beta \tilde \epsilon_i}/Z_F$ yields
\begin{equation}
\label{Gaudin}
 (e^{\beta \tilde
     \epsilon_i}\pm 1)
\langle \tilde Ac_i \rangle_{gc}= \langle [c_i, \tilde A]_\pm\rangle_{gc}~.
\end{equation}
For $\tilde A=c_j^\dagger$ one obtains the expected result
\begin{equation}
 \langle  c_j^\dagger
c_i\rangle_{gc}=\delta_{ij}\frac{1}{e^{\beta \tilde
     \epsilon_i}\pm 1}= \delta_{ij}\langle n_i \rangle_{gc}
\end{equation}
For the case where
$\tilde A$ is a
product of $m$ creation and $m-1$ annihilation operators
we return to the fermionic case with $i\to l_m$ and 
 $\tilde A=P^{(m-1)} c_{l_m}^\dagger$ addressed in
 Eq. (\ref{Wickproduct}).
As all $l_i$ in $P^{(m)}$
  differ, $c_{l_m}$ commutes with $  P^{(m-1)}$ i.e.
$[c_{l{m}}, P^{(m-1)}c^\dagger_{l{m}}]_+= P^{(m-1)} $
 leading to 
\begin {equation}
\langle n_{l_1}... n_{l_m}\rangle_{gc}
=\langle n_{l_1}... n_{l_{m-1}}\rangle_{gc}\langle n_{l_m}\rangle_{gc}~.
\end {equation}
Iteration leads to the complete factorization.
Despite the fact that the direct derivation of Eq. (\ref{Wickproduct})
using the factorization of $\rho_{gc}^{(0)}$ is much simpler,
 Gaudin's method was shown, as an extension of it
 can be used also for the case of the canonical ensemble.

\subsection{The canonical ensemble}

As Eq. (\ref{productform}) also holds in the canonical ensemble
one has again only to address expectation values of products of
occupation number operators. In reference 3 a general expression 
for the expectation value of $m$-particle operators in the position
and spin representation was presented. Their Eq. (2) leads
for the expectation value of a product of $m$ different occupation number
operators to
\begin{equation}
\label{TKKT1}
\langle n_{l_1}... n_{l_m}\rangle_c=
\sum_{k=m}^N(-1)^{k-m}Z_{N-k}~s^{(k)}_{l_1l_2...l_m} 
\end{equation}
with
\begin{equation}
\label{TKKT2}
s^{(k)}_{l_1l_2...l_m}=\sum_{j_1=1}^k\sum_{j_2=1}^k...
\sum_{j_m=1}^ke^{-\beta\sum_{i=1}^m j_i\epsilon_{l_i}}\delta_{j_1+j_2+...+j_m,k}~.
\end{equation}
 This generalizes the expressions for $m=1$ and $m=2$
presented in section II. For the case of bosons the factor $(-1)^{k-m} $
is missing. We return to this expression in appendix A. 

In the following we propose a new approach to the calculation of 
$\langle n_{l_1}... n_{l_m}\rangle_c$ which provides analytical
expressions in the limiting cases for the models discussed in section
IV.\\

In the canonical ensemble the cyclic move of $c_i$ in the trace 
in Eq. (\ref{cyclic})
is not allowed as $c_{l_m}|~\rangle_N$ leaves the Hilbert space with
fixed $N$. We therefore have to proceed differently here.
We treat expectation values of the type 
 $ \langle \bar Ac_j^\dagger c_i \rangle$, where $j\ne i$ and $\bar A$ is an
 arbitrary product of $m-1$ creation and $m-1$ annihilation operators.
The cyclic move of
 $c_j^\dagger  c_i$ in the trace  in Eq. (\ref{cyclic}) is possible
 in the grand canonical as well as the canonical ensemble.
Therefore
no index for the expectation values is used in the following.
The relation 
\begin{equation}
e^{\beta H_0} c^\dagger_jc_i e^{-\beta  H_0}
=e^{\beta (\epsilon_j- \epsilon_i)}  c^\dagger_jc_i 
\end{equation}
holds without and with the tilde on $H_0$. In order to obtain an 
equation for  $ \langle \bar Ac_j^\dagger c_i \rangle$ we here use
$ c_j^\dagger c_i\bar A=  \bar Ac_j^\dagger c_i+[c_j^\dagger c_i, \bar
A]$,  with the commutator for fermions as well as bosons
 after the cyclic move. This leads to  
\begin{equation}
\label{Gaudin2}
(e^{\beta (\epsilon_i- \epsilon_j)}-1)  \langle \bar Ac_j^\dagger c_i
\rangle = \langle [c_j^\dagger c_i, \bar A]\rangle~.
\end{equation}
In order to solve this equation for $ \langle \bar Ac_j^\dagger c_i
\rangle $ the one-particle energies of the states $j$ and $i$
have to {\it differ}. We later discuss this condition in more detail
and assume $ \epsilon_i \ne \epsilon_j$ in the following.
The choice  $\bar A=c_i^\dagger c_j$ leads to
 the simplest
nontrivial relation. This
gives a formula for $ \langle c_i^\dagger c_j c_j^\dagger c_i \rangle
=\langle n_i(1 \mp n_j)\rangle$. With $[c_j^\dagger c_i, c_i^\dagger
c_j]=n_j-n_i$ Eq. (\ref{Gaudin2}) leads to 
\begin{equation}
\label{simple}
 \langle c_i^\dagger c_j c_j^\dagger c_i \rangle=
\frac{\langle n_j\rangle-\langle n_i\rangle}{e^{\beta (\epsilon_i- \epsilon_j)}-1}~, 
\end{equation}
valid for both ensembles and fermions as well as bosons. A detailed 
discussion of this result in  a slightly modified form will be presented
later.

In the
following we focus on $ \langle P^{(m)} \rangle$ defined in
Eq. (\ref{Wickproduct}).
 Only if the spectrum of
one-particle energies $\epsilon_i $ is non-degenerate a complete
treatment is possible as all quantum numbers $l_i$ in the product
$P^{(m)}$  differ. This is e.g. the case for one-dimensional 
spinless fermions in an external potential, like a box potential
or a harmonic well 
treated as an example in the next section. 

Using the anticommutation rule we rewrite
the last two occupation number operators in $P^{(m)}$ 
in the spirit of the simple example just discussed
\begin {equation}
n_{l_{m-1}}n_{l_m}=n_{l_{m-1}}-c^\dagger_{l_{m-1}}c_{l_m} c^\dagger_{l_m}c_{l_{m-1}}~.
\end {equation}
With
 \begin {equation}
\bar A=  P^{(m-2)}c^\dagger_{l_{m-1}}c_{l_m} 
\end {equation} 
the product of
the occupation number operators is given by
\begin {equation}
\label{Pn}
P^{(m)}= P^{(m-1)}-\bar A c^\dagger_{l_m}c_{l_{m-1}}
\end {equation}
The commutator in Eq. (\ref{Gaudin2}) is readily evaluated 
as $c^\dagger_{l_m}c_{l_{m-1}}$ commutes with $P^{(m-2)}$. With 
$[ c^\dagger_{l_m}c_{l_{m-1}},c^\dagger_{l_{m-1}}c_{l_m}]=n_{l_m}-n_{l_{m-1}}$  
used earlier one obtains 
\begin{equation}
 \langle [c^\dagger_{l_m}c_{l_{m-1}} , \bar A] \rangle=
\langle P^{(m-2)}(n_{l_m}-n_{l_{m-1}})\rangle~.
\end{equation}
From Eqs. (\ref{Pn}) and (\ref{Gaudin2}) one obtains in the
non-degenerate case assumed in the following
\begin{equation}
\label{nrecursive}
\langle P^{(m)}\rangle =
 \frac{\langle  P^{(m-2)}n_{l_{m-1}}\rangle e^{\beta
     \epsilon_{l_{m-1}}}
-\langle  P^{(m-2)}n_{l_m}\rangle e^{\beta
     \epsilon_{l_m}}}{ e^{\beta \epsilon_{l_{m-1}}}-e^{\beta\epsilon_{l_m}}}~.
\end{equation}
This holds for the canonical as well as the grand canonical averages.
For the case of the canonical ensemble this  
 relation could have been found earlier by using 
Eqs. (\ref{TKKT1}) and  (\ref{TKKT2}) (see appendix A).

For $m=2$ this equation reads
\begin{equation}
\label{nn}
\langle n_{l_1} n_{l_2}\rangle= 
\frac{\langle n_{l_1}\rangle e^{\beta \epsilon_{l_1}}
-\langle n_{l_2}\rangle e^{\beta
  \epsilon_{l_2}}}{e^{\beta\epsilon_{l_1}}-
e^{\beta \epsilon_{l_2}}}~.
\end{equation}
This is a slightly different version of Eq. (\ref{simple}) which holds for
fermions as well as bosons.
One obtains a minus sign in front of the expression on the rhs of 
 Eq. (\ref{nn}) for the case of bosons.

As a test for the grand canonical ensemble
one can put in the Fermi functions
for the $\langle n_{l_i}\rangle $ and readily sees the factorization which holds
in contrast to the canonical ensemble. The deviations from the factorization
in this case will be
quantitatively studied for simple models in the next section. 

With the result for $m=2$ the calculation of  $\langle P^{(3)}\rangle$
using Eq. (\ref{nrecursive}) suggests the following general result
\begin{equation}
\label{generalresult}
\langle n_{l_1}....n_{l_m}\rangle 
=\sum_{i=1}^m  \langle n_{l_i}\rangle\prod_{j(\neq i)}^m
\frac{e^{\beta\epsilon_{l_i}}}{e^{\beta\epsilon_{l_i}}-e^{\beta\epsilon_{l_j}}}~.
\end{equation}
It is obviously fullfilled for $m=2$. If this is inserted on 
the right hand side of  Eq. (\ref{nrecursive})
for the expectation values of $m-1$ occupation number operators,
simple algebra presented in Appendix B  
completes the inductive proof of Eq. (\ref{generalresult}).
Together with Eq. (\ref{productform}) this shows that one can 
express the expectation value of an arbitrary $m$-particle operator
in terms of the expectation values $\langle n_{l_i}\rangle$
also within the canonical ensemble. 
Obviously this new result is
more complicated than Wick's theorem Eq. (\ref{Wickproduct})
for the grand canonical average. The complete factorization in this case 
easily follows from Eq. (\ref{generalresult}). This is also shown in
Appendix B.

From the fact that the use of the Fermi functions in Eq. (\ref{generalresult})
leads to the factorization of the expectation value 
one expects that the deviations from Wick's
theorem in the canonical ensemble are large for quantum
numbers $l_i$ where the $\langle n_{l_i}\rangle_c$ differ
strongly from  $\langle n_{l_i}\rangle_{gc}$.
To test this quantitatively we calculate  the ``Wick ratio''
\begin{equation}
\label{rW}
r_W^{l_1l_2}(T,N)\equiv \frac{\langle n_{l_1} n_{l_2}\rangle_c}
{\langle n_{l_1}\rangle_c 
\langle n_{l_2} \rangle_c}
\end{equation}
 as well as the correspondingly defined Wick ratio 
for higher products of occupation number operators for special models .
The deviation from Wick's theorem is quantified by how much $r_W$ differs
from one.

\section{Applications}

In this section we present quantitative results for the deviation
from Wick's theorem in the canonical ensemble for two rather different
models.

 Fermions in one dimension in a harmonic potential are an
example of a system with an equidistant one-particle spectrum. Such a system
can be realized in ultracold gases \cite{Jochim}.
 In the non-interacting case exact
results for the thermodynamic properties and the mean occupation
numbers can be obtained with a recursive method \cite{KS1}.
Here we use these results for  
the mean occupation numbers in Eqs. (\ref{nn}) and (\ref{generalresult}) to 
calculate expectation values of products of occupation number
operators for arbitrarily large numbers $N$ of fermions. This model
also plays an important role in the context of the Tomonaga-Luttinger
model \cite{Haldane}.

In order to address the problem of Eqs. (\ref{nn}) and
(\ref{generalresult}) with degeneracies of the one-particle energies
a simple ``semiconductor model'' is studied with zero width 
of the valence
and conduction bands. The particle number $N$ is chosen to be equal
to the number of valence band states.
For this simple model a direct combinatorical method can be used
 to calculate the
canonical mean occupation numbers. It avoids the numerical problems
when using Eq. (\ref{nsum}) and easily allows to understand how the
grand canonical results arises in the large $N$ limit.

\subsection{Fermions in a one-dimensional harmonic trap}

We consider $N$ noninteracting spinless fermions
in a system with
 nondegenerate  equidistant  one-particle 
energies $\epsilon_i$  
\begin{equation}
 \epsilon_i=i\Delta~, ~~~i=1,2,3,....,\infty~.
\end{equation}
 By performing the sum over $n_1$ in 
Eq. (\ref{pf}) only, as a first step,
a recursive approach leads to an explicit analytical
expression for $Z_N$ \cite{KS1}. This canonical partition
function has the form as for a system of $N$
uncoupled harmonic oscillators with frequencies
$\omega_m=m\Delta/\hbar, ~m=1,...,N$.
Proceeding similarly for the mean occupation numbers one obtains
the recursion relation \cite{KS1}
\begin{equation}
\langle n_i \rangle_N =e^{-\beta N\Delta}\langle n_{i-1} \rangle_N
+(1-e^{-\beta N\Delta}) \langle n_{i-1} \rangle_{N-1}.
\end{equation}
Together with the ``Schmidt relation'' Eq. (\ref{nrecursion}) one can obtain a
recursion relation between the mean occupation numbers with the {\it same}
total particle number only
\begin{equation}
\label{KSrecursion}
\langle n_{i+1} \rangle_N =1- e^{-\beta N\Delta}
-( e^{-\beta (N-i)\Delta}- e^{-\beta N\Delta}) \langle n_i \rangle_N~.
\end{equation}
Using it with $\langle n_1 \rangle_N=1-e^{-\beta N\Delta} $
as the starting point, provides an ``upward'' way to
calculate the $\langle n_i \rangle_N$. Alternatively one can use
Eq. (\ref{KSrecursion}) to express  $\langle n_i \rangle_N$   in terms
of $\langle n_{i+1} \rangle_N$ and start the downward 
iteration for $m+1\gg N$
with  $\langle n_{m+1} \rangle_N=0$.  Together this provides an
efficient numerically stable procedure to calculate the 
mean occupation numbers.
In the general case one has to compare $k_BT$ with {\it two}
energy scales, $\Delta$ and  $N\Delta$. In the scaling limit
$N\to \infty$ with fixed $\beta \Delta$ the recursion relation
Eq. (\ref{KSrecursion}) for $\bar n_l\equiv  
\langle n_{N+l} \rangle_N$ 
simplifies to 
\begin{equation}
\label{recursionscaling}
\bar n_{l+1}=1-q^{-l}\bar n_l~,~~~~\bar n_l=q^l(1-\bar n_{l+1}),
\end{equation}
whith $q=e^{-\beta \Delta}$. 
For $1d$ fermions with a linear dispersion this scaling limit
corresponds to the addition of an infinite ``Dirac sea''\cite{SM}.
Due to the symmetry relation \cite{KS1}
\begin{equation}
\bar n_{-l}=1-\bar n_{l+1}
\end{equation}
only the upward or the downward recursion has to be used.
As long as $k_BT \ll N\Delta$ holds, Eq. (\ref{recursionscaling})
provides an excellent approximation for very large but finite $N$.
Only how $k_BT$  compares to $\Delta$ matters in this limit.
For $(N\Delta \gg) k_BT \gg \Delta$ the $\bar n_l$ approach the 
grand canonical Fermi function $f_l=1/(e^{\beta(l-1/2)\Delta}+1)$,
while for $k_BT\ll \Delta$ there are appreciable deviations \cite{KS1}.\\

After this summary of previous results we address the Wick ratio $r_W$
for this model. With the definition  $\bar n_l\equiv  
\langle n_{N+l} \rangle_N$ 
 Eq. (\ref{nn}) reads 
\begin{equation}
\label{nnequi}
\langle n_{N+i+l}n_{N+i}\rangle=
\frac{\bar n_{i+l}-q^l \bar n_i}{1-q^l}
\end{equation}
For arbitrary values of $i$ and $l$ the expectation value 
$\langle n_{N+i+l}n_{N+i}\rangle$ follows from the numerical values for
the mean occupation numbers. As in the scaling limit $N\to \infty$
various analytical results can be obtained we focuss on this limit where
Eq. (\ref{recursionscaling}) can be used to obtain the mean occupation
numbers.

\begin{figure}
\label{figho}
\centering
\epsfig{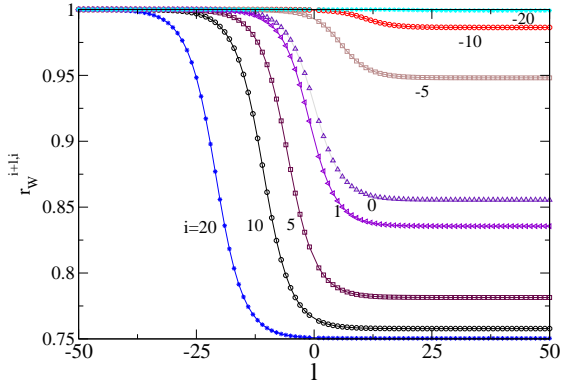}
\vspace{0.2cm}
\caption{Wick ratio $r_W^{i+l,i}$ for $q=e^{-\beta \Delta}=0.75$ as a
 function of $l(\ne 0)$ for
different values of $i$ indicated at the corresponding ``curves''. 
The asymtotic values for large positive and negative values
of $l$ are discussed in the text.  } 
\end{figure}
 In Fig. 1 we present results for the Wick ratio
$r_W^{i+l,i}= \langle n_{N+i+l}n_{N+i}\rangle/(\bar n_{i+l}\bar n_i)$
for $q=0.75$ as a function of $l(\ne 0)$ for different values of $i$.
The asymptotic values for $l \to \pm \infty$ can be understood
analytically using Eq. (\ref{recursionscaling}). For large enough $l$
the second equation implies $\bar n_{i+l}\approx q^{i+l}$ and
\begin{equation}
\bar n_{i+l}-q^l\bar n_i \approx q^{i+l}(1-q^{-i}\bar n_i)=
q^{i+l}\bar n_{i+1}\approx \bar n_{i+l} \bar n_{i+1} ~.
\end{equation}
 This implies 
\begin{equation}
l \to \infty:~~ r_W^{i+l,i}\to      \bar n_{i+1}/\bar n_i ~.
\end{equation}
 The
asymptotic values for large $l$ in Fig. 1 agree with this analytical result.
 For large values of $i$
the  ratio $ \bar n_{i+1}/\bar n_i  $ tends to $q$.

For $l\to -\infty$ Eq.(\ref{nnequi}) implies
$\langle n_{N+i+l}n_{N+i}\rangle \to \bar n_i $ and $\bar n_{i+l}\to 1  $
i.e. $ r_W^{i+l,i}\to 1  $ in agreement with Fig. 1. 
We finally present results for the expectation value of the triple
product $P_i^{(3)}\equiv n_{N+i},n_{N+i+1},n_{N+i+2}$.
It determines the probability for three neighbouring
one-particle levels to be all occupied. 
 Using
Eq. (\ref{generalresult}) it is given by
\begin{equation}
\label{drei1}
\langle P_i^{(3)} \rangle_c=\frac{1}{(1-q)(1-q^2)}\left [
\bar n_i q^3-\bar n_{i+1}(q+q^2)+ \bar n_{i+2}\right ]~.
\end{equation}
As $\bar n_i\to 1$ for $i \to -\infty$ 
one obtains $ \langle P_i^{(3)} \rangle_c \to 1 $
and the  Wick ratio $r_W^{i,i+1,i+2}
\equiv \langle P_i^{(3)} \rangle_c/(\bar n_i\bar n_{i+1}\bar n_{i+2}) $
 tends to one in agreement with
the numerical results of Fig. 2.

For the discussion of the limit $i \to \infty$ Eq. (\ref{drei1})
cannot directly be used.
Applying the second form of Eq. (\ref{recursionscaling}) twice,
 Eq. (\ref{drei1}) can be rewritten in a form which allows to discuss
this limit
\begin{equation}
\label  {drei2}
\langle P_i^{(3)} \rangle_c=\frac{q^iq^{i+1}q^3}{(1-q)(1-q^2)}\left [
\bar n_{i+2}-\bar n_{i+3}(1+q)+ \bar n_{i+4}q \right ]~.
\end{equation}
For $i\gg 1$ and $j=1,2,3$ the expectation values  $\bar n_{i+j}$
are again using  Eq. (\ref{recursionscaling})  to  a sufficient
approximation given by $q^{i+j}$. This yields $\langle P_i^{(3)} \rangle_c
\approx q^iq^{i+1}q^{i+2}  q^3$, i.e.
\begin{equation}
\langle  n_{N+i,N+i+1,N+i+2}\rangle_c \approx \bar n_i \bar n_{i+1}
\bar n_{i+2}~q^3~.
\end{equation}
For large $i$ the Wick ratio $r_W^{i,i+1,i+2}$ is therefore given
by $q^3$ in agreement with the numerical results in
 Fig. 2 for $q=0.5$ and $0.75$.  For $q=0.9$ one has to go larger values of 
$i$ to see the asymptotic behaviour.

\begin{figure}
\centering
\epsfig{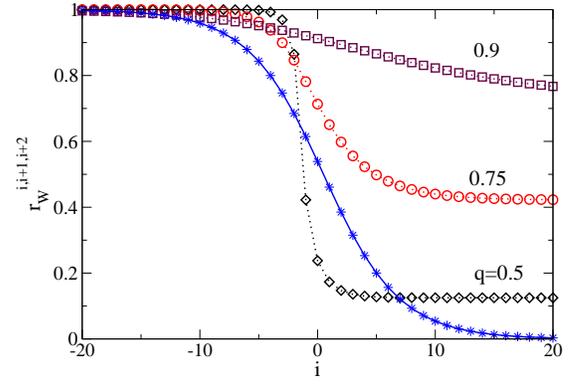}
\vspace{0.2cm}
\caption{Wick ratio $r_W^{i,i+1,i+2}$ for $q=e^{-\beta \Delta}=0.5, 0.75$
and $0.9$ as a
 function of $i$. 
The asymtotic values for large positive and negative values
of $i$ are discussed in the text. The blue ``curve'' (stars)
shows $\bar n_i$ for $q=0.75$  } 
\end{figure}
Realistic values of $q$ differ for the two applications of this model mentioned
earlier.
For fermions in a 1d harmonic trap the value of the temperature and $\Delta$
can be independently experimentally tuned, i.e. $q$ can be chosen
 quite arbitrarily.
For free fermions  with a linearized dispersion 
$\Delta \sim 1/L$ holds for  box of length $L$
 and the limit $L\to \infty$ leads to $q\to 1$,
implying a Wick ratio of one as in the grand canonical ensemble.\\

\subsection{Zero bandwidth semiconductor model}

A simple model with $M$ degenerate valence band states and 
 $M$ degenerate conduction band states is considered
\begin{equation}
 H=\sum_{i=1}^M[\epsilon_vc_{v,i}^\dagger c_{v,i}+
(\epsilon_v+\Delta)c_{c,i}^\dagger c_{c,i}]=\epsilon_v{\cal N}_v+
(\epsilon_v+\Delta){\cal N}_c~.
\end{equation}
In the following we put $\epsilon_v=0$.
Despite the fact that the general case $N\ne M$ is as easily treated
as the special case $N=M$, we only present results for
 the latter in the following.
In this case the $N$-fermion ground state is given by the filled
valence band. The excited states have $m$
holes in the valence band and $m$ particles in the conduction band. 
The number of ways to chose $m$ holes in the $N$ valence band states is 
given by $\binom{N}{m}$.
 The same result is obtained for the number
of ways to put the $m$ particles in the conduction band. Therefore
the canonical partition function is given by
\begin{equation}
Z_N=\sum_{m=0}^N\binom{N}{m}^2e^{-m\beta\Delta}~.
\end{equation}  
In order to obtain the mean occupation numbers $\langle
n_{\alpha,i}\rangle_c$ for $\alpha=v,c$ it is sufficient to calculate
$\langle {\cal N}_c\rangle_c$ as   $\langle n_{\alpha,i}\rangle_c$ does
not depend on $i$ and   $\langle  {\cal N}_v\rangle_c +
\langle  {\cal N}_c \rangle_c=N$ holds.
Introducing the probability distribution $p_N(m)$ for the number 
of electrons in the conduction band
 the mean occupation
of the conduction band is given by
\begin{equation}
\label{pNm}
\langle {\cal N}_c\rangle_c=\sum_{m=0}^Nmp_N(m)~; 
~~~ p_N(m)=\frac{1}{Z_N}\binom{N}{m}^2e^{-m\beta\Delta}~.
\end{equation}  
For not too small values of $q=e^{-\beta\Delta}$ the probability 
distribution $p_N(m)$ for large $N$ resembles a Gaussian distribution
as shown in Fig. 3 for $N=50$. For $k_BT/\Delta=0.5$ also the grand
 canonical result is shown
 for comparison (filled squares) . While its average value
is close to the canonical result its width is 
significantly larger. This is discussed quantitatively later.

\begin{figure}
\label{fighc}
\centering
\epsfig{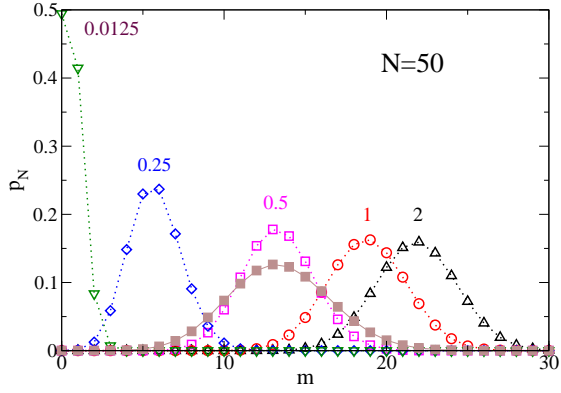}
\vspace{0.2cm}
\caption{Probability distribution $p_N(m)$ for the number $m$
of conduction electrons for the
zero bandwidth semiconductor model for $N=50$ and different values of 
$k_BT/\Delta$ indicated at the maximum of the corresponding curve.
For $k_BT/\Delta=0.5$   also the  grand canonical probability
distribution is shown (filled squares)} 
\end{figure}

 In the grand canonical ensemble the chemical potential
lies  in the
 middle of the bands, $\mu=\Delta/2$, for the case $N=M$
considered here. This guarantees 
$\langle {\cal N}\rangle_{gc}=N$ for all temperatures
\begin{equation}
f_c\equiv \langle n_{c,i} \rangle_{gc}=\frac{1}{e^{\beta \Delta/2}+1}=
1-\langle n_{v,i} \rangle_{gc}~.
\end{equation}
 This is
in contrast to the general case $N\ne M$, in which the chemical potential
is temperature dependent. Due to the factorization of
$\rho_{gc}^{(0)}$ the conduction band states $c,i$
are independently empty with probability $1-f_c$ and filled
with probability $f_c$. Therefore the grand canonical
distribution function is binomial
\begin{equation}
p_{gc}(m)= \binom{N}{m}f_c^m(1-f_c)^{N-m}~.
\end{equation} 
with average value $Nf_c$ and mean square deviation $Nf_c(1-f_c)$.   
  The transition of a 
binomial distribution to a Gaussian distribution in the large $N$-limit
 discussed in
textbooks on statistical mechanics.
Before addressing the
width of $p_N(m)$ the average occupation numbers  $\langle n_{c,i}
\rangle_c$ and $f_c$ are compared.\\

 In Fig. 4 we show results from the numerical evaluation of
$ \langle n_{c,i} \rangle_c $ using Eq. (\ref{pNm}).
\begin{figure}
\label{fighc}
\centering
\epsfig{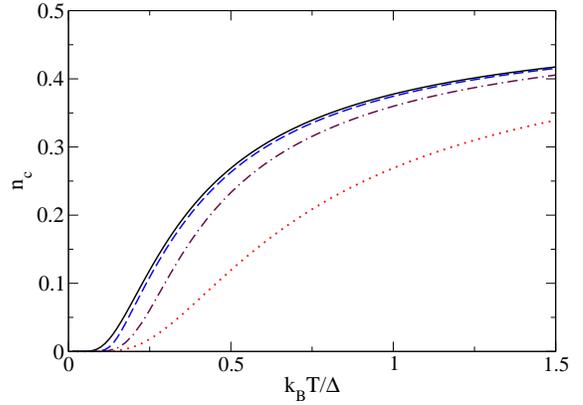}
\vspace{0.2cm}
\caption{Mean occupation numbers $n_c=\langle n_{c,i}\rangle_c$ for the
zero bandwidth semiconductor model as a function of temperature for different
values of $N(=M)$: $N=1$ (dotted), $N=4$ (dashed-dotted) and $N=20$
(dashed).  The grand 
canonical result is presented by the full curve} 
\end{figure}
For small values of $N$ the deviations of the canonical from the
grand canonical result are rather
large.
In the high temperature limit the grand canonical result
is approached. 

   In the extreme low temperature limit
 $N^2e^{-\beta \Delta} \ll 1$ one approximately has 
$\langle n_{c,i}\rangle\approx Ne^{-\beta \Delta}$ which 
deviates strongly from  the grand canonical result $\approx e^{-\beta \Delta/2}$.
 
In the large $N$ limit
$\langle {\cal N}_c\rangle \approx \bar m$ where $\bar m$ is the
position of the maximum of  $p_N(m)$. Using the  Stirling formula
 $m!\approx \sqrt{2\pi m}(m/e)^m$ 
valid for $m\gg 1$ the variable $m$ in  $p_N(m)$
can be treated as continuous and 
 the position $\bar m$ of the maximum of $p_N(m)$ is easily obtained
setting the derivative of $p_N(m)$ or $ \ln(p_N(m))$ to zero. 
With $d~\ln m!/dm\approx \ln m$ one obtains
\begin{eqnarray}
\frac{d~\ln(p_N(m))}{dm}&\approx& 2~[-\ln m+\ln (N-m)]+\ln q
 \nonumber \\
&=&2 ~\ln\left
(\frac{N-m}{m}q^{1/2} \right )~. 
\end{eqnarray}  
Putting the argument of the logarithm equal to $1$ the position of the
maximum follows as $\bar m =N/(q^{-1/2}+1)$. With $q=e^{-\beta
  \Delta}$ and $\langle n_{c,i} \rangle =\langle {\cal N}_c\rangle/N$
one finally obtains in the large $N$ limit
\begin{equation}
\label{largeN}
\langle n_{c,i} \rangle_c \to \frac{1}{e^{\beta \Delta/2}+1}=
\langle n_{c,i} \rangle_{gc}~.
\end{equation}

Next we address the expectation values $\langle n_{\alpha,i}
n_{\beta,j}\rangle_c$ with $i \ne j$ 
when $\alpha=\beta$, in the canonical ensemble.
 As they are independent of $i$ and $j$ there are
only three different types: the $\langle n_{\alpha,i}
n_{\alpha,j}\rangle_c$ with $\alpha=v$ or $c$ and 
 $\langle n_{c,i} n_{v,j}\rangle_c$. The
value of the latter follows easily from Eq. (\ref{nn}).
As shown in the following the $\langle n_{\alpha,i}
n_{\alpha,j}\rangle_c$ are determined by  $\langle n_{c,i}
n_{v,j}\rangle_c$ and $ \langle n_{\alpha,i} \rangle_c $. This stems from
the fact that  $\langle {\cal N}_c {\cal N}_v \rangle_c$ 
 can be expressed in terms of
 $\langle {\cal N}_\alpha^2\rangle_c$ and  $\langle {\cal N}_\alpha\rangle_c$
\begin{equation}
\label{quadratic}
\langle  {\cal N}_c {\cal N}_v \rangle_c=\sum_{m=0}^Nm(N-m)p_N(m)=
N\langle {\cal N}_c\rangle_c-\langle {\cal N}_c^2\rangle_c
\end{equation}
and with $m(N-m)=N(N-m)-(N-m)^2$ the index $c$ on the right hand
side can be replaced by $v$.
Using $\langle  {\cal N}_c {\cal N}_v \rangle_c=N^2 \langle n_{c,i}
n_{v,j}\rangle_c$ this implies
\begin{equation}
\langle {\cal N}_\alpha^2\rangle_c= N^2(\langle n_{\alpha,i}\rangle_c- 
 \langle n_{c,i} n_{v,j}\rangle_c)~.
\end{equation}
 For $i\ne j$ one has
 $\langle {\cal
  N}_\alpha^2\rangle_c= N(N-1)\langle n_{\alpha,i}n_{\alpha,j}\rangle_c
+N\langle n_{\alpha,i}\rangle_c $. This leads to the promised result
\begin{equation}
\label{ncnc}
\langle n_{\alpha,i}n_{\alpha,j}\rangle_c=\langle n_{\alpha,i}\rangle_c   
-\frac{N}{N-1} \langle n_{c,i} n_{v,j}\rangle_c ~.
\end{equation}
This allows to calculate the $\langle
n_{\alpha,i}n_{\alpha,j}\rangle_c $ in terms of $ \langle n_{\alpha,i}\rangle_c$   
and $\langle n_{c,i} n_{v,j}\rangle_c$ 
 given by Eq. (\ref{nn})
\begin{equation}
\label{nnHalb}
\langle n_{c,i} n_{v,j}\rangle_c=\frac{ \langle n_{c,i}\rangle_c
-e^{-\beta \Delta}(1- \langle n_{c,i}\rangle_c)}{1-e^{-\beta \Delta}}~.
\end{equation} 
If $ \langle n_{c,i} n_{v,j}\rangle_c  $ factorizes in the limit
$N\to \infty$ Eq. (\ref{ncnc}) implies the same for the
$\langle n_{\alpha,i}n_{\alpha,j}\rangle_c $ for $i\ne j$.\\

In Fig. 5 we show the Wick ratios $r_W^{vv},r_W^{cv}$ and $r_W^{cc}$ as a
 function of temperature for $N=4$ and $N=20$. The limiting
values for $T\to 0$ and $T \to \infty$ can be understood analytically. At $T=0$
the valence band is completely occupied, i.e. $\langle n_{v,i}\rangle_c=1
=\langle n_{v,i} n_{v,j}\rangle_c$. This implies
\begin{equation}
r_W^{vv}(T=0,N)=1~.
\end{equation}
 As at zero temperature 
the conduction band is empty  $\langle n_{c,i}\rangle_c=0 $,
$\langle n_{c,i} n_{v,j}\rangle_c =0$ and $\langle n_{c,i} n_{c,j}\rangle_c=0 $
holds. For the Wick ratios $r_W^{cv}$ and  $r_W^{cc}$
 one therefore encounters a $"0/0"$ problem
and the limit $T\to 0$ has to be studied. 
\begin{figure}
\centering
\epsfig{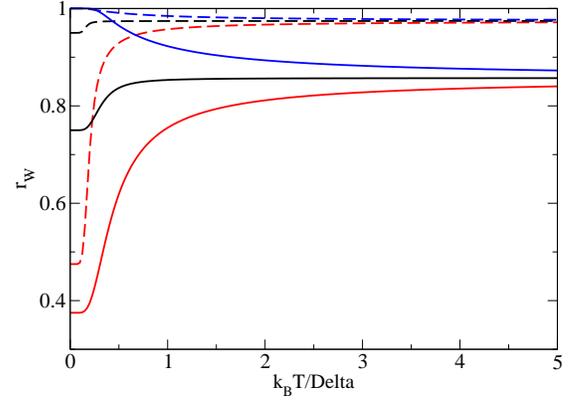}
\vspace{0.2cm}
\caption{Wick ratios $r_W^{vv}$ (blue curves),  $r_W^{cv}$ (black curves)
and $r_w^{cc}$ (red curves) as a function of $k_BT/\Delta$
for  $N=4$ (full curves) and $20$ (dashed curves).} 
\end{figure}
 As mentioned earlier, in the extreme low temperature limit 
$\langle n_{c,i}\rangle_c\approx Ne^{-\beta \Delta}\ll 1$ holds.
 With Eq. (\ref{nnHalb})
this leads to 
\begin{equation}
\label{lowT}
r_W^{cv}(T\to 0,N)= 1-\frac{1}{N}~.
\end{equation}
Alternatively this can be
obtained by simple combinatorics. As the state $v,j$
is supposed to be occupied  there are $N-1$ ways to promote a valence
electron to the state $c,i$, leading to  
 $\langle n_{c,i} n_{v,j}\rangle_c\approx (N-1)e^{-\beta \Delta}$.
With $\langle n_{c,i}\rangle_c(1-\langle n_{c,i}\rangle_c)\approx
\langle n_{c,i}\rangle_c\approx  Ne^{-\beta \Delta} $ this leads to the result in
Eq. (\ref{lowT}).
We finally address the $T \to 0$ limit of $r_W^{cc}$.  
 There are $N(N-1)/2$ ways to put two electrons
into two prescribed conduction band states, leading to  
 $\langle n_{c,i} n_{c,j}\rangle_c\approx N(N-1)e^{-2\beta \Delta}/2$.
With the result for  $\langle n_{c,i}\rangle_c$ one obtains
\begin{equation}
\label{lowT2}
r_W^{cc}(T\to 0,N)= \frac{1}{2}\left
  (1-\frac{1}{N} \right )~.
\end{equation}

 In the high temperature limit $T \to \infty$
simply counting numbers of states determines 
$\langle n_{\alpha,i} n_{\beta,j}\rangle_c $ with $\alpha,i$
differing from $\beta,j$.
The number of ways to put two fermions in these two one-particle states
 and the other $N-2$ particles into the remaining
$2N-2$ states is given by $\binom{2N-2}{N-2}$ and the partition function
by the total number of possible states $\binom{2N}{N}$. This leads to
\begin{equation}
\label{highT}
T \to \infty:~~~~r_W \to \frac{1-1/N}{1-1/(2N)}
\end{equation}
independently of the upper indices.
 Without the
combinatorics just presented, this value of $r_W$ can be obtained
from Eq. (\ref{ncnc}) realizing that the $\langle
n_{\alpha,i}n_{\beta,j}\rangle_c$ with $\alpha,i\ne \beta,j$ are all the same in
the infinite temperature limit. One can solve this
equation for $x=\langle n_{c,i} n_{v,j}\rangle_c=
\langle n_{\alpha,i} n_{\alpha,j}\rangle_c$ using $\langle
n_{\alpha,i}\rangle_c=1/2$ and  obtains the result of Eq. (\ref{highT}).\\

In Fig. 6 we show the Wick ratio $r_W^{cv}$  as a function of $1/N$
for different values of $k_BT/\Delta$. The results lie between the 
``curves'' determined by Eqs. (\ref{lowT}) and (\ref{highT}). For large
$N$  the infinite temperature result is reached quickly with increasing
temperature.

\begin{figure}
\centering
\epsfig{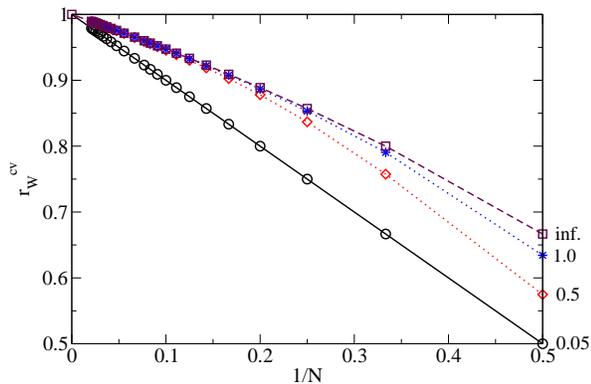}
\vspace{0.2cm}
\caption{Wick ratio $r_W^{cv}$  as a function of $1/N$
for different values of $k_BT/\Delta$, indicated at the right edge of the 
figure. The lower and upper boundary curve
 of the results are discussed in the text. } 
\end{figure}

We now return to the comparison of the widths of $p_N$ and $p_{gc}$ 
both shown for $k_BT/\Delta=0.5$ in Fig. 3. The mean square deviation
$\sigma_c^2$ for both cases is  given by
\begin{eqnarray}
 \sigma_c^2=\langle {\cal N}_c^2 \rangle &-&\langle {\cal N}_c \rangle^2
= N^2(\langle n_{c,1}n_{c,2}\rangle -\langle n_{c,1}\rangle^2)
\nonumber \\
&+& N(\langle n_{c,1}\rangle-\langle n_{c,1}n_{c,2}\rangle)~.
\end{eqnarray}
Using Wick's theorem for
 the grand canonical case the first term on the right hand side
vanishes, leading to
$(\sigma_c^2)_{gc}=Nf_c(1-f_c)$, as mentioned earlier.

For the canonical ensemble $\sigma_c^2$ can be expressed in terms of 
$ \langle n_{c,i}\rangle_c$ and $r_W^{cc}$ as
\begin{equation}
\label{sigma2c}
 (\sigma_c^2)_c=N^2\langle n_{c,1}\rangle_c^2(r_W^{cc}-1)+
N(\langle n_{c,1}\rangle_c-r_W^{cc}\langle n_{c,1}\rangle_c^2)~.
\end{equation}
For large $N$ and not too low temperatures $r_W^{cc}-1\approx -1/(2N)$
holds and the first term of $\sigma_c^2 $ which vanishes in the grand
canonical case is approximately given by $-N\langle
n_{c1}\rangle_c^2/2$. To leading order in $N$ the second term is
 given by $N \langle n_{c1}\rangle_c(1-\langle n_{c1}\rangle_c)
\approx Nf_c(1-f_c)$. In the high temperature limit $(\sigma_c^2)_{gc}\approx
2 (\sigma_c^2)_c$ holds, i.e. the width of $p_{gc}$ is larger by a
factor $\sqrt{2}$ than that of $p_N$ as can be confirmed in Fig. 3.

\section{Summary}

With an approach similar to Gaudin's proof of Wick's theorem 
for the grand canonical ensemble new results for expectation 
values of products of occupation numbers
of one-particle states with differing one-particle energies
were presented for noninteracting fermions
in Eqs. (\ref{nn}) and (\ref{generalresult}).
They are valid for the grand canonical as well as the canonical
ensemble. To arbitrary order of the products
the expectation values
are expressed in terms of the average occupation numbers. For two
different models it was explicitely shown that these relations
allow a deeper undestanding of the deviations from Wick's theorem
in the canonical ensemble which go beyond the purely numerical approaches
presented earlier. The deviations can be very large at low
temperatures if the product involves occupation number operators
of one-particle states which are unocupied at zero temperature.

\section{Acknowledgements}

The author wants to thank V. Meden and W. Zwerger
 for a critical reading of the manuscript
and useful comments.

\begin{appendix}

\section{Alternative derivation of Eq. (\ref{nn})}

Here we show how Eq. (\ref{nrecursive})  for the canonical ensemble
 could have been found 
 using Eqs. (\ref{TKKT1})  and (\ref{TKKT2}).

 In Eq. (\ref{TKKT2} )
the sums in $s_{l_1...l_m}^{(k)}$ run from $1$ to $k$. As $j_i\ge 1$
for all $i$ it is obvious that the largest value a $j_i$  can take
is $k-m+1$. 
As the  upper limit
of the sums one can also take $\infty$, as the Kronecker delta does its
job. Therefore in the following the upper limits of the sums are suppressed.  

If one multiplies $s_{l_1...l_m}^{(k)}$ by $e^{\beta \epsilon_m}$ this leads
after changing the summation index $j_m$ by one to 
\begin{eqnarray}
\label{sk}
 e^{\beta \epsilon_m} s^{(k)}_{l_1...l_m}&=&\sum_{j_1=1}\sum_{j_2=1}...
\sum_{j_m=0}e^{-\beta\sum_{i=1}^{m} j_i\epsilon_{l_i}}\delta_{j_1+...+j_m,k-1} \nonumber \\
&=& s^{(k-1)}_{l_1...l_{m-1}}+ s^{(k-1)}_{l_1..l_{m-1}l_m}~.
\end{eqnarray}
In taking the difference with the according expression where one multplies
with $e^{\beta \epsilon_{m-1}}$, the second terms cancel and one obtains
\begin{equation}
\label{sk2}
(e^{\beta \epsilon_m}-e^{\beta \epsilon_{m-1}}) s^{(k)}_{l_1...l_m}=
 s^{(k-1)}_{l_1...l_{m-1}}- s^{(k-1)}_{l_1...l_{m-2}l_m}~.
\end{equation}
Eq. (\ref{sk}) reads for $m\to m-1$
\begin{equation}
 e^{\beta \epsilon_{m-1}} s^{(k)}_{l_1...l_{m-1}}= s^{(k-1)}_{l_1...l_{m-2}}+ 
s^{(k-1)}_{l_1..l_{m-2}l_{m-1}}~.
\end{equation}
If one performs the corresponding multiplication with  $e^{\beta
  \epsilon_m}$ and takes the difference the comparison with
Eq. (\ref{sk2}) yields
\begin{equation}
(e^{\beta \epsilon_m}-e^{\beta \epsilon_{m-1}}) s^{(k)}_{l_1...l_m}=
 e^{\beta \epsilon_{m-1}} s^{(k)}_{l_1...l_{m-1}}- 
e^{\beta \epsilon_{m}} s^{(k)}_{l_1...l_{m-2}l_m}~.
\end{equation}
Inserting this into Eq. (\ref{TKKT1}) leads to 
\begin{eqnarray}
(e^{\beta \epsilon_{m-1}}-e^{\beta \epsilon_m})\langle
n_{l_1...l_m}\rangle_c&=& \nonumber \\
 \langle n_{l_1...l_{m-1}}\rangle_c e^{\beta \epsilon_{m-1}}
&-& \langle n_{l_1...l_{m-2}l_m} \rangle_c e^{\beta \epsilon_m}.
\end{eqnarray}
For $\epsilon_{m-1}\ne \epsilon_m$  division proves
Eq. (\ref{nrecursive}) in a way different from the one presented in 
section III.

\section {Induction step in the proof of Eq. (31)}

In this appendix the inductive step in the proof of
Eq. (\ref{generalresult})
 is presented.
Using the   abbreviation $x_i= e^{\beta \epsilon_{l_i}}$ we assume the formula
\begin{equation}
\label{generalapp}
\langle n_{l_1}....n_{l_{m-1}}\rangle 
=\sum_{i=1}^{m-1}  \langle n_{l_i}\rangle\prod_{j(\neq i)}^{m-1}
\frac{x_i}{x_i-x_j}
\end{equation}
to be correct. By putting this into the
recursion formula Eq. (\ref{nrecursive}) 
\begin{equation}
\label{recursiveapp}
\langle P^{(m)}\rangle =
 \frac{\langle  P^{(m-2)}n_{l_{m-1}}\rangle x_{m-1}
-\langle  P^{(m-2)}n_{l_m}\rangle x_m} 
{x_{m-1}-x_m}~.
\end{equation}
we show that formula Eq. (\ref{generalapp}) also holds for 
$m$. In both
$\langle P^{(m-2)}n_{l_k}\rangle$ with $k=m-1$ and $k=m$ all occupation number
operators $n_{l_j}$ with $j\le m-2$ occur. In contrast  $n_{l_{m-1}}$
and  $n_{l_ m}$ only appear in one of the two expectation values in
 Eq (\ref {recursiveapp}  ).
A term proportional to  $ \langle n_{l_{m-1}}  \rangle$ only results from the 
first expectation value. Its contribution to $\langle P^{(m)}\rangle$
is given by
\begin{equation}
\frac{x_{m-1}}{x_{m-1}-x_m}\prod_{j=1}^{m-2}\frac{x_{m-1} }
{x_{m-1}-x_j  }=
\prod_{j(\neq m-1)}^{m}\frac{x_{m-1} }
{x_{m-1}-x_j  }~.
\end{equation}
The term proportional to  $ \langle n_{l_{m}}  \rangle$ results from the second
expectation value
\begin{equation}
-\frac{x_{m}}{x_{m-1}-x_m}\prod_{j=1}^{m-2}\frac{x_m }
{x_m-x_j  }=
\prod_{j(\neq m)}^{m}\frac{x_m }
{x_m-x_j  }~.
\end{equation}
For the terms proportional to  $\langle n_{l_j} \rangle$ with $j\le m-2$
it is sufficient to consider a single example e.g.  $\langle n_{l_1}
\rangle$.
With $p_{m-2}^{(1)}=\prod_{j=2}^{m-2}x_1/(x_1-x_j)$
 and using the recursion relation Eq. (\ref{recursiveapp}) the
contribution proportional to  $ \langle n_{l_1}  \rangle$  in  
 $\langle P^{(m)}\rangle$ is given by 
\begin{eqnarray}
&~&\frac{1}{x_{m-1}-x_m}p_{m-2}^{(1)}\left (
\frac{x_1x_{m-1}}{x_1-x_{m-1}}-\frac{x_1x_m}{x_1-x_m}\right) \nonumber
\\
&=& \prod_{j=2}^m\frac{x_1}{x_1-x_j}~.
\end{eqnarray}
This completes the inductive step for the proof of Eq. 
(\ref{generalresult}).

The proof that $\langle P^{(m)} \rangle$ completely factorizes for the 
grand canonical average is again easier by induction. For $m=2$
one has
\begin{eqnarray}
\langle n_{l_1}n_{l_2} \rangle_{gc}&=&
\frac{x_1\langle n_{l_1}\rangle_{gc}-x_2\langle n_{l_2}\rangle_{gc}}
{x_1-x_2} \\
&=& \frac{x_1(x_2+1)-x_2(x_1+1)}{(x_1-x_2)(x_1+1)(x_2+1)}
=\langle n_{l_1} \rangle_{gc}\langle n_{l_2} \rangle_{gc} \nonumber
\end{eqnarray}
Now we assume that $\langle P^{(m-1)} \rangle_{gc}$ completely factorizes
and use Eq. (\ref{recursiveapp}) for the induction step. This
assumption implies $\langle P^{(m-2)}n_{l_k} \rangle_{gc}$ 
with $k=m-1$ and $k=m$ factorizes  $\langle P^{(m-2)}n_{l_k} \rangle_{gc}
=\langle P^{(m-2)} \rangle_{gc} \langle n_{l_k} \rangle_{gc}$.
This implies with  Eq. (\ref{recursiveapp})
\begin{eqnarray}
\langle n_{l_1}....n_{l_m}\rangle_{gc}&=& \langle P^{(m-2)} \rangle_{gc}
\frac{x_{m-1}  \langle n_{l_{m-1}} \rangle_{gc}- 
x_{m-1}  \langle n_{l_{m-1}} \rangle_{gc}}{x_{m-1}-x_m} \nonumber \\
&=&\prod_{i=1}^m\langle n_{l_i}\rangle_{gc}~.
\end{eqnarray}
This proof is certainly more involved than the one in subsection IIIb.   

\end{appendix}

\end{document}